\documentclass[%
reprint,
superscriptaddress,
showpacs,
 amsmath,amssymb,
 aps,
prb,
]{revtex4-1}

\usepackage{dcolumn}
\usepackage{bm}

\usepackage[dvipdfmx]{graphicx,color}

\usepackage{ulem} 

\begin{document}

\preprint{APS/123-QED}

\title{Density-matrix renormalization group study of third harmonic generation in one-dimensional Mott insulator coupled with phonon} 

\author{Shigetoshi Sota}
\email{sotas@riken.jp}

\affiliation{%
Computational Materials Science Research Team,
RIKEN Advanced Institute for Computational Science (AICS), Kobe, Hyogo 650-0047, Japan
}%

\author{Seiji Yunoki}
\affiliation{%
Computational Materials Science Research Team,
RIKEN Advanced Institute for Computational Science (AICS), Kobe, Hyogo 650-0047, Japan
}%
\affiliation{%
Computational Condensed Matter Physics Laboratory, RIKEN, Wako, Saitama 351-0198, Japan
}%
\affiliation{%
Computational Quantum Matter Research Team,
RIKEN Center for Emergent Matter Science (CEMS), Wako, Saitama 351-0198, Japan
}%

\author{Takami Tohyama}
\email{tohyama@rs.tus.ac.jp}
\affiliation{
Department of Applied Physics, Tokyo University of Science, Tokyo 125-8585, Japan
}%

\date{\today}

\begin{abstract}
We examine the third-order non-linear optical response of a one-dimensional Mott insulator coupled with phonons. The Mott insulator is described by an extended Hubbard-Holstein model. The third harmonic generation (THG) of the model is calculated by dynamical density-matrix renormalization group. We find that the electron-phonon interaction enhances the intensity of THG if the on-site Coulomb interaction is comparable to the band width, but the enhancement is small for a realistic parameter value of a typical Mott insulator Sr$_2$CuO$_3$. The low-energy spin excitation does not contribute to the optical response if the electron-phonon coupling is neglected. We find that the THG spectrum can detect the spin excitations even for no electron-phonon coupling. The introduction of the electron-phonon coupling leads to a slight increase of low-energy weight in THG without peak structure through phonon-assisted spin excitation. To fully understand the phonon-assisted spin excitation, we additionally calculate the linear susceptibility of a doped Hubbard-Holstein model, showing a spectral distribution different from that at half filling.

\end{abstract}

\pacs{78.20.Bh, 78.47.N-}


\maketitle


\section{Introduction}

In one-dimensional (1D) Mott insulators, photoexcitation across the Mott gap creates two particles, i.e., an unoccupied site called holon and a doubly occupied site of electrons called doublon. They can move inside the system without being disturbed by surrounding spins in the background.  This is a manifestation of a separation of the charge and spin degrees of freedom, called the spin-charge separation inherent in 1D correlated electron systems.~\cite{Maekawa} Charge excitation in 1D Mott insulators is therefore insensitive to the spin degree of freedom. Once electron-phonon (EP) interaction is taken into account in the 1D Mott insulators, the spin degree of freedom also participates in the charge excitation. The optical conductivity has clearly shown the presence of phonon-assisted spin excitation inside the Mott gap in a typical 1D Mott insulator Sr$_2$CuO$_3$.~\cite{Suzuura, Lorenzana} A dynamical density-matrix renormalization group (DMRG) calculation has clearly shown that a 1D extended Hubbard-Holstein model containing Holstein-type coupling of electron to the Einstein phonons can explain both the Mott-gap excitation and phonon-assisted spin excitation in the optical conductivity.~\cite{Sota}

The linear susceptibility $\chi^{(1)}$ with respect to the applied electric field, which is related to the optical conductivity, provides information on the dipole-allowed odd-parity states among the photoexcited states.  The third-order nonlinear optical susceptibility $\chi^{(3)}$ is useful to detect not only the odd-parity states but also the dipole-forbidden states with even parity.~\cite{Butcher}  The analyses of $\chi^{(3)}$ in Sr$_2$CuO$_3$ have suggested that odd- and even-parity states are nearly degenerate with a large transition dipole moment between them.~\cite{Kishida1,Ogasawara} Theoretically, $\chi^{(3)}$ in the 1D Mott insulators has been examined by employing the numerically exact diagonalization technique for small clusters of the Hubbard model at half filling.~\cite{Mizuno}  It has been shown that odd- and even-parity states are almost degenerate in the same energy region and that the degeneracy is due to the spin-charge separation and strong on-site Coulomb interaction.~\cite{Tohyama}  This situation is described by an effective model with one hole and one doubloon called  holon-doublon model.~\cite{Mizuno} The holon-doublon model reproduces very well the characteristic behaviors of the experimental $\chi^{(3)}$ including data from the third-harmonic generation (THG) spectroscopy.~\cite{Kishida2}

The THG spectrum obtained by the holon-doublon model does not show low-energy excitation below the three-photon absorption peak.~\cite{Takahashi} The holon-doublon model completely ignores the spin degree of freedom. It is thus interesting to clarify whether non-linear charge excitation due to spin degrees of freedom appears in a full spin model like a Hubbard model. Furthermore, The presence of the EP interaction in Sr$_2$CuO$_3$ is crucial for the understanding of phonon-assisted spin excitations in $\chi^{(1)}$ as mentioned above. Therefore, the effect of the EP interaction not only on $\chi^{(1)}$ but also on $\chi^{(3)}$ is necessary to be clarified on an equal footing.

In this paper we examine $\chi^{(3)}$ obtained by THG for the 1D Hubbard-Holstein model at half filling. We employ the dynamical DMRG.~\cite{Sota} We find that the EP interaction increases the intensity of three-photon absorption in THG if the on-site Coulomb interaction is comparable to the band width, but such an increase is small for a realistic parameter set of Sr$_2$CuO$_3$. We find a low-energy spectral weight in the extended Hubbard model, which is absent in the holon-doublon model. The origin of the excitation is attributed to the spin degree of freedom. The introduction of the EP interaction leads to a slight increase of low-energy weight in THG without a peak structure. To fully understand the phonon-assisted spin excitation, we additionally calculate $\chi^{(1)}$ of a single-hole doped Hubbard-Holstein model. The spectral distribution near the energy of the photo-assisted spin excitation is found to be strongly modified by the presence of the single hole. 

The rest of this paper is organized as follows.  We introduce the extended Hubbard-Holstein model and show outlines of the procedure to calculate $\chi^{(1)}$ and $\chi^{(3)}$ in Sec.~\ref{sec2}.  In Sec.~\ref{sec3}, calculated results of the THG are presented. The effects of the EP interaction of THG are discussed for two parameter sets of the model. The low-energy spectral weight associated with the spin degree of freedom appears in the extended Hubbard model without phonons. The EP interaction induces a slight enhancement of intensity. In order to show a delicate nature of the phonon-assisted spin excitation, we demonstrate $\chi^{(1)}$ for the extended Hubbard-Holstein model away from half filling. The data exhibits a spectral distribution different from that at half filling, indicating the sensitivity of the phonon-assisted spin excitation.  The summary is given in Sec.~\ref{sec4}.

\section{\label{sec2}Model and Method}
We consider a 1D extended Hubbard-Holstein model at half filling that has been used in the previous work of the optical conductivity.~\cite{Sota} The Hamiltonian is defined by 
\begin{eqnarray}
H &=& -t\sum_{i,\sigma}(c_{i,\sigma}^{\dagger}c_{i+1,\sigma}+\mathrm{H.c.})
 + U\sum_{i}n_{i,\uparrow}n_{i,\downarrow} \nonumber \\
&& + V\sum_{i}(n_{i}-1)(n_{i+1}-1) \nonumber \\
&& + \omega_{0}\sum_{i}b_{i+1/2}^{\dagger}b_{i+1/2}  \nonumber \\
&& -g\sum_{i}(b_{i+1/2}^{\dagger}+b_{i+1/2})(n_{i}-n_{i+1}),  
\label{H}
\end{eqnarray}
where $c_{i,\sigma}^{\dagger}$ ($c_{i,\sigma}$) is the creation (annihilation) operator of an electron at site $i$ with spin $\sigma$, and $b_{i+1/2}^{\dagger}$ ($b_{i+1/2}$) is the creation (annihilation) operator of a phonon at site $i+1/2$ keeping in mind oxygen sites in cuprates. This model includes electron hopping, $t$, on-site and nearest-neighbor Coulomb repulsions, $U$ and $V$, respectively, phonon frequency, $\omega_{0}$, and EP coupling, $g$.

The linear susceptibility $\chi^{(1)}$ in 1D systems is given by
\begin{equation}
\chi^{(1)} (-\omega; \omega)=\frac{1}{\epsilon_0 L} \frac{e^2}{\hbar}
\sum_a \left ( \frac{x_{0a} x_{a0}}{\Omega_a-i\delta_a-\omega} + \frac{x_{0a} x_{a0}}{\Omega_a+i\delta_a+\omega} \right ) ,
\label{chi1}
\end{equation}
where $L$ is the number of sites, $\epsilon_0$ is the dielectric constant, $ex_{0a}$ is the dipole moment between the ground state $|0 \rangle$ and excited state $|a \rangle$ with odd parity, $\Omega_a$ is the energy difference between $|0 \rangle $ and $|a \rangle$, and $\delta_a$ is the damping factor.  The third-order susceptibility $\chi^{(3)}$ is expressed as
\begin{eqnarray}
&&\chi^{(3)} (-\omega_\sigma; \omega_1, \omega_2, \omega_3) \nonumber \\
&&=\frac{1}{\epsilon_0 L} \frac{e^4}{3!\hbar^3} {\bf {\cal P}}\times \nonumber \\
&&\sum_{a,b,c} \frac{x_{0a} x_{ab} x_{bc} x_{c0}}{(\Omega_a-i\delta_a-\omega_\sigma)(\Omega_b-i\delta_b-\omega_2-\omega_3)(\Omega_c-i\delta_c-\omega_3)},\label{chi3}
\end{eqnarray}
where $b$ and $c$ denote even and odd states, respectively, and ${\bf {\cal P}}$ represents the sum of permutation on $\omega_1$, $\omega_2$, $\omega_3$, and $\omega_\sigma=\omega_1+\omega_2+\omega_3$.  Hereafter, we take $e=\hbar=\epsilon_0=1$.  The damping factors, $\delta_a$, $\delta_b$ and $\delta_c$, are assumed to have the same infinitesimal value, $\delta$, for all excited states of the systems.

The THG spectrum is proportional to $|\chi^{(3)}(-3\omega;\omega,\omega,\omega)|$. The dominant contribution to THG comes from 
\begin{eqnarray}
&&\chi^{(3)}_0(-3\omega;\omega,\omega,\omega) \nonumber \\
&&=\frac{1}{L} \sum_{a,b,c} \frac{x_{0a} x_{ab} x_{bc} x_{c0}}{(\Omega_a-i\delta-3\omega)(\Omega_b-i\delta-2\omega)(\Omega_c-i\delta-\omega)}.
\label{THG}
\end{eqnarray}
In our calculations, we take only this term for simplicity. The contribution of other terms in Eq.~(\ref{chi3}) is expected to be small.~\cite{Takahashi} 

By using the dipole operator $\hat{x}$, the equation~(\ref{THG}) can be rewritten as
\begin{eqnarray}
&&\chi^{(3)}_0(-3\omega;\omega,\omega,\omega) \nonumber \\
&&=\frac{1}{L}\left\langle 0 \right| \hat{x} \frac{1}{H-3\omega-i\delta} \hat{x} \frac{1}{H-2\omega-i\delta} \hat{x} \frac{1}{H-\omega-i\delta} \hat{x} \left| 0\right\rangle.
\label{THG3}
\end{eqnarray}
In dynamical DMRG calculations, we need to prepare target states appropriate to the corresponding quantities. In Eq.~(\ref{THG3}), we target seven states for a given energy $\omega$: (i) $\left| 0\right\rangle$, (ii) $\hat{x}\left| 0\right\rangle$, (iii) $\left( H-\omega-i\delta \right)^{-1}\hat{x}\left| 0\right\rangle$, (iv) $\hat{x} \left( H-\omega-i\delta \right)^{-1}\hat{x}\left| 0\right\rangle$, (v) $\left( H-3\omega-i\delta \right)^{-1}\hat{x}\left| 0\right\rangle$, (vi) $\hat{x} \left( H-3\omega-i\delta \right)^{-1}\hat{x}\left| 0\right\rangle$, and (vii) $\left( H-2\omega-i\delta \right)^{-1}\hat{x}\left( H-\omega-i\delta \right)^{-1}\hat{x}\left| 0\right\rangle$. The (iii), (iv), (v), and (vi) target states are evaluated by using a kernel-polynomial expansion method given in Ref.~4. The (vii) target state is obtained by performing the polynomial expansion twice. The final form of Eq.~(\ref{THG3}) is given by the product of the (vi) and (vii) target states.

In our calculations, we use a 24-site chain that is the same as the calculation of optical conductivity of Sr$_2$CuO$_3$.~\cite{Sota} The parameters related to phonons are fixed to $g/t=0.4$ and $\omega_0/t=0.25$ that are realistic values for cuprates. The number of phonons is taken to be five per every site. In our kernel-polynomial expansion method, the Lorentzian broadening $\delta$ is replaced by a Gaussian broadening with half width at half maximum 0.2$t$. We use the truncation number $m$ in the DMRG process to be $m=2000$ and a truncation error is less than $10^{-3}$.

\section{\label{sec3}Results}
\begin{figure}[t]
\includegraphics[width=8.0cm]{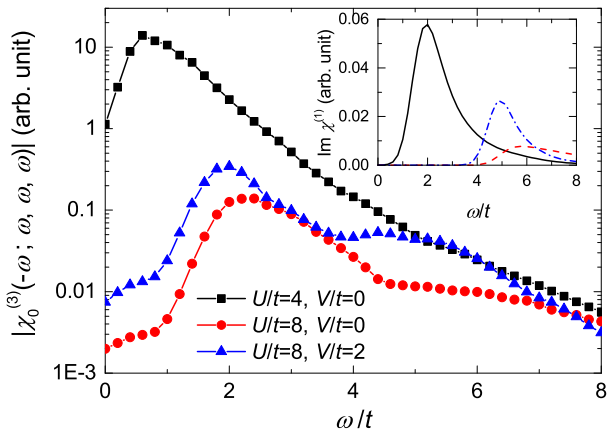}
\caption{(Color Online)
$|\chi^{(3)}_0(-3\omega;\omega,\omega,\omega)|$ of a 24-site half-filled extended Hubbard chain. The squares, triangles, and circles represent the result of $(U/t, V/t)=$(4,0), (8,0), and (8,2), respectively. The inset shows $\mathrm{Im}\chi^{(1)} (-\omega; \omega)$ for $(U/t, V/t)=$(4,0) (solid line), (8,0) (broken line), and (8,2) (dotted-broken line). 
}
\label{fig1}
\end{figure}

Before looking at the effect of the EP interaction, let us examine the effect of $U$ and $V$ on THG. Figure~\ref{fig1}(a) shows $|\chi^{(3)}_0(-3\omega;\omega,\omega,\omega)|$ for the extended-Hubbard model together with $\mathrm{Im}\chi^{(1)}(-\omega; \omega)$ (see inset). For $U/t=8$ and $V/t=0$, there is a peak at $\omega/t=2.2$, which is due to three-photon absorption according to the peak position of $\mathrm{Im}\chi^{(1)}$. A broad hump structure around $\omega/t=3$ may come from two-photon absorption. A broad structure centered at $\omega/t=6$ is due to single-photon absorption. A small enhancement of intensity around $\omega/t=0.5$ may be related to low-energy excitations due to spin degrees of freedom. Note that there is no linear response in this energy region. We also emphasize that in the holon-doublon model, where the spin degrees of freedom is neglected, this enhancement does not appear.~\cite{Takahashi}

Adding $V/t=2$ to the model with $U/t=8$, we find an enhancement of the THG intensity (see triangles in Fig.~\ref{fig1}). The enhancement has the same origin as the increase of $\mathrm{Im}\chi^{(1)}$, which is due to the formation of an exciton. Keeping $V/t=0$ and reducing $U/t$ down to the band width, i.e., $U=4t$, we also find a strong enhancement of the THG intensity as expected from the enhancement of $\mathrm{Im}\chi^{(1)}$. 

\begin{figure}[t]
\includegraphics[width=8.0cm]{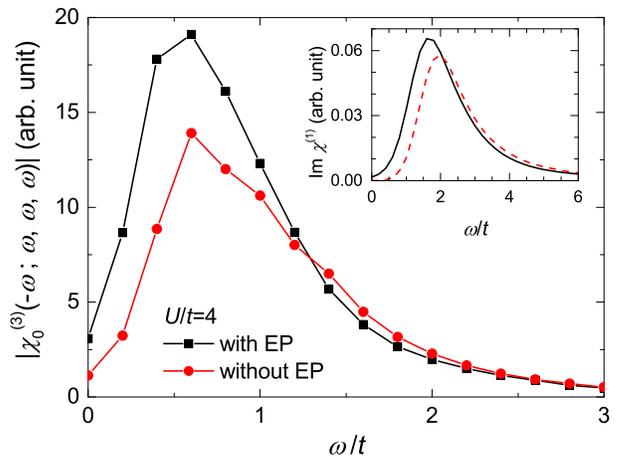}
\caption{(Color Online)
$|\chi^{(3)}_0(-3\omega;\omega,\omega,\omega)|$ of a 24-site half-filled Hubbard-Holstein chain with $U/t=4$. The squares and circles represent the result with and without the EP interaction, respectively. $g/t=0.4$ and $\omega_0/t=0.25$. The inset shows $\mathrm{Im}\chi^{(1)} (-\omega; \omega)$ with (solid line) and without (broken line) the EP interaction.
}
\label{fig2}
\end{figure}

\begin{figure}[t]
\includegraphics[width=8.0cm]{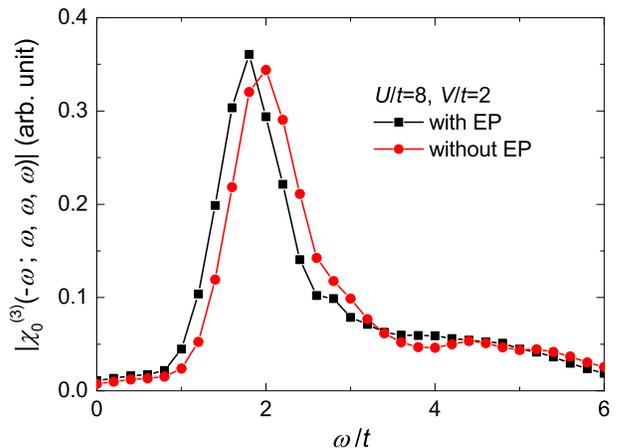}
\caption{(Color Online)
$|\chi^{(3)}_0(-3\omega;\omega,\omega,\omega)|$ of a 24-site half-filled extended Hubbard-Holstein chain with $U/t=8$ and $V/t=2$. The squares and circles represent the result with and without the EP interaction, respectively. $g/t=0.4$ and $\omega_0/t=0.25$.
}
\label{fig3}
\end{figure}

Figure~\ref{fig2} shows the effect of the EP interaction on THG for $U/t=4$. Since the EP interaction effectively reduces $U$, the THG intensity as well as $\mathrm{Im}\chi^{(1)}$ (see inset) increases with the interaction. Figure~\ref{fig3} shows the effect of the EP interaction for the case of $U/t=8$ and $V/t=2$ realistic for Sr$_2$CuO$_3$. The peak position slightly shifts to lower energy, though the peak intensity little changes. Comparing with Fig.~\ref{fig2}, we conclude that the effect of the EP interaction on the intensity is thus larger for smaller value of $U$.

The low-energy excitation near $\omega/t=0.5$ in Fig.~\ref{fig3} that is originated from low-lying spin excitation slightly increases. The phonon-assisted spin excitation seen in $\mathrm{Im}\chi^{(1)}$ (Ref.~4) may contribute to the increase, but its effect is not so significant to observe a hump structure unlike the case of $\mathrm{Im}\chi^{(1)}$. This would indicate that detecting the phonon-assisted spin excitation is sensitive to optical processes.

\begin{figure}[t]
\includegraphics[width=8.0cm]{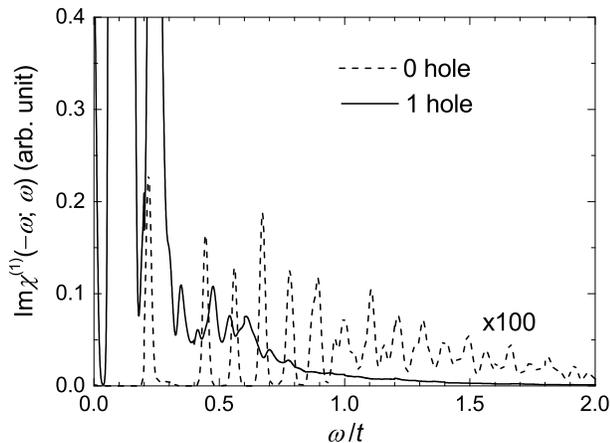}
\caption{
$\mathrm{Im}\chi^{(1)}(-\omega; \omega)$ of a 24-site extended Hubbard-Holstein chain with $U/t=8$ and $V/t=2$ in the energy region below the Mott gap. The broken line represents phonon-assisted spin excitations at half filling, whose intensity is multiplied by 100. The solid line represents a case away from half filling, where the number of up (down) spin is 12 (11). $g/t=0.4$, $\omega_0/t=0.25$, and a Gaussian broadening with half width at half maximum 0.01$t$.
}
\label{fig4}
\end{figure}

In order to demonstrate such sensitivity, we demonstrate the effect of a mobile carrier on the phonon-assisted spin excitation. Figure~\ref{fig4} exhibits the low-energy spectrum of $\mathrm{Im}\chi^{(1)}$ for the 24-site extended Hubbard-Holstein model. At half filling (the broken line), many excitations with small intensity appear above the phonon frequency $\omega_0/t=0.25$, being assigned to the phonon-assisted spin excitations.~\cite{Suzuura,Lorenzana,Sota} With removing a down-spin electron from the system (twelve up-spin electrons and eleven down-spin electrons), we find a large peak at $\omega/t=0.2$ (see the solid line), which is attributed to a metallic excitation. Around $\omega/t=0.25$, there is a structure with strong intensity that comes from phonon excitations. The phonon intensity is nearly 300 times larger than that for half filling, since a mobile carrier can induce phonons around itself. Above the phonon structure, we find a broad but small intensity distributed up to higher-energy. The broad structure is expected to contain phonon-assisted spin excitation as is the case at half filling. 
However, the distribution of spectral weight is different from that for the half-filled case as shown by broken line. Therefore, it is clear that the presence of a mobile hole modifies strongly the spectral distribution of the phonon-assisted spin excitation.

\section{\label{sec4}Summary}
We have examine the third-order nonlinear optical response obtained by THG for the 1D extended Hubbard-Holstein model at half filling. We have used the dynamical DMRG to calculate the spectra. We have found that the EP interaction increases the intensity of three-photon absorption in THG if the on-site Coulomb interaction is comparable to the band width, but such an increase is small for a realistic parameter set of Sr$_2$CuO$_3$. We have also found that there is a low-energy spectral weight in the extended Hubbard model below the three-photon absorption peak. Such a weight is absent in the holon-doublon model where the spin degree of freedom is ignored. Thus, the origin of the excitation is attributed to the spin degree of freedom. The introduction of the EP interaction leads to a slight increase of low-energy weight in THG without a peak structure unlike linear optical absorption. To fully understand the phonon-assisted spin excitation, we have additionally calculated the linear optical response of a single-hole doped Hubbard-Holstein model. The spectral distribution near the energy of the photo-assisted spin excitation is found to be strongly modified by the presence of the single hole, indicating the sensitivity of the phonon-assisted spin excitation.

\begin{acknowledgments}
We acknowledge H. Okamoto and H. Kishida for useful and stimulating discussions. 
This work is financially supported by MEXT HPCI Strategic Programs for Innovative Research (SPIRE) (hp120283, hp130007, hp140215) and Computational Materials Science Initiative (CMSI). Numerical calculation was partly carried out at the K computer, the RIKEN Advanced Institute for Computational Science, and the Supercomputer Center, Institute for Solid State Physics, University of Tokyo. This work was also supported by Grant-in-Aid for Scientific Research (No. 26287079 and No. 22740225) from MEXT, Japan.
\end{acknowledgments}

\nocite{*}



\end{document}